\documentclass[12pt]{article}

\usepackage{amssymb}

 \csname
@addtoreset\endcsname{equation}{section}

\def\a{\alpha}
\def\b{\beta}
\def\c{\gamma}
\def\C{\Gamma}
\def\d{\delta}

\def\e{\epsilon}

\def\f{\phi}
\def\F{\Phi}

\def\k{\kappa}

\def\L{\Lambda}
\def\m{\mu}

\def\s{\sigma}

\def\t{\tau}
\def\th{\theta}

\def\x{\xi}\def\y{\eta}
\def\bp{\bar{\pi}}

\def\O{\Omega}
\def\o{\omega}

\def\una{\underline{\alpha}}

\newcommand{\scs}[1]{\section{{\sc #1}}}
\newcommand{\scss}[1]{\subsection{{\sc #1}}}

\def\cF{{\cal F}}

\def\cN{{\cal N}}
\def\cR{{\cal R}}

\def\cV{{\cal V}}

\def\superphi{\mbox{{\boldmath $\phi$}}}

\def\yb{{\bar y}}
\def\zb{{\bar z}}
\def\pb{{\bar \pi}}
\def\Ah{{\widehat A}}
\def\Fh{{\widehat \F}}
\def\hA{{\widehat A}}
\def\hF{{\widehat \F}}
\def\ad{\dot{\a}}
\def\bd{\dot{\b}}
\def\cd{\dot{\c}}

\def\nn{\nonumber}
\def\fr{\frac}
\def\ra{\rightarrow}

\def\be{\begin{eqnarray}}
\def\ee{\end{eqnarray}}
\def\ba{\begin{array}}
\def\ea{\end{array}}
\def\bec{\begin{center}}
\def\ec{\end{center}}
\def\ns{\normalsize}
\def\ft#1#2{{\textstyle{{\scriptstyle #1} \over {\scriptstyle #2}}}}
\def\mx#1#2#3#4{\left#1\begin{array}{#2} #3 \end{array}\right#4}
\def\la#1{\label{#1}}
\def\eq#1{(\ref{#1})}
\let\bm=\bibitem

\newcommand{\hoch}[1]{$^{#1}$}

\begin{document}

\newpage
\hfill{{MIFP-05/02}}
\\[-20pt]

\hfill{{UUITP-16/02}}
\\[-20pt]

\hfill{hep-th/0211113}

\vspace{20pt}

\begin{center}


{\Large\bf\sc Superspace Formulation of \\[10pt] 4D Higher Spin Gauge
Theory}


\vspace{30pt}

{\large J. Engquist\hoch1, E. Sezgin\hoch2 and
P. Sundell\hoch3}\\[20pt]

\be \phantom{aaaa}\ba{rl} \mbox{\small \it
\hoch{1,3}}&\!\!\!\!\mbox{\small \it Department of Theoretical
Physics, Uppsala
University,} \\&\!\!\!\!\mbox{\small\it Box 803, 751 08 Uppsala, Sweden}\\[5pt]
\mbox{\small\it\hoch2}&\!\!\!\!\mbox{\small \it George P. and
Cynthia W. Mitchell Institute for Fundamental Physics,}\\
&\!\!\!\!\mbox{\small \it Texas A\&M University, College Station,
TX 77843-4242, USA}\ea\nn\ee

\vspace{100pt}

{\large\bf Abstract}

\end{center}

Interacting AdS$_4$ higher spin gauge theories with $\cN\geq 1$
supersymmetry so far have been formulated as constrained systems
of differential forms living in a twistor extension of 4D
spacetime. Here we formulate the minimal $\cN=1$ theory in
superspace, leaving the internal twistor space intact. Remarkably,
the superspace constraints have the same form as those defining
the theory in ordinary spacetime. This construction generalizes
straightforwardly to higher spin gauge theories $\cN\geq 2$
supersymmetry.

\newpage


\scs{Introduction}


The $\cN=8$ massless higher spin (HS) gauge theory in $D=4$
\cite{Vasiliev:1991vu} is expected to arise as a consistent
truncation of M theory on $AdS_4\times S^7$ in an unbroken phase
describing the theory at high energies \cite{us3,Sezgin:2002rt}.
Various $\cN=1,2,4$ HS gauge theories can arise in less symmetric
compactifications of M theory \cite{Engquist:2002vr}. In this
paper we shall only discuss massless HS gauge theories theories,
though in general massive fields are needed for embedding in M
theory. From the 3d holographic point of view, the unbroken phase
of M theory on $AdS_4\times S^7$ has been conjectured to
correspond to an $SU(N)$ invariant free singleton theory
\cite{us3,Sezgin:2002rt}, and the massless sector to an $O(N^2-1)$
vector model \cite{Klebanov:2002ja}. Similar ideas in the context
of the Type IIB theory on $AdS_5\times S^5$ have been discussed in
\cite{su2,us1,Vasiliev:2001zy,edseminar,Sezgin:2002rt}.

Interacting AdS$_4$ HS gauge theories with $\cN\geq 1$
supersymmetry so far have been formulated as constrained systems
of differential $0$-forms and $1$-forms living in the product of
4D spacetime with an internal twistor space
\cite{Vasiliev:1991vu}. In this formulation the local HS
symmetries are realized as internal gauge symmetries, and
spacetime diffeomorphisms are incorporated into the gauge group as
local field dependent translations. By expanding in curvatures
\cite{Vasiliev:1990yr} it is possible to obtain the non-linear
field equations in a manifestly reparametrization invariant form
\cite{Sezgin:2002ru}. For a better understanding of the
geometrical structures underlying HS gauge theories it would be
desirable to extend the diffeomorphism symmetry as to include
additional generators of the HS symmetry algebra. A natural first
step in this direction is to reformulate the HS gauge theories in
superspace with manifest superdiffeomorphism invariance. The
superspace formulation of free 4D conformal HS gauge theories with
$\cN\geq 1$ supersymmetry has been given in
\cite{Vasiliev:2001zy}. The superspace formulation of the
linearized AdS$_4$ HS gauge theory based on a particular $\cN=2$
algebra has been given in \cite{Gates:1997xs}.

In this paper we extend the integrable systems describing
interacting AdS$_4$ HS gauge theories, by replacing the 4D
spacetime by a $D=(4|4\cN)$ superspace with $4\cN$ anti-commuting
$\th$-coordinates. This introduces extra spinorial directions in
the $1$-forms as well as $\th$-dependence in all component fields.
On the other hand, there are also new constraints coming from
projections of the differential form constraints in the new
spinorial directions. As a result, each supermultiplet in the
spectrum is described by a single constrained superfield, and we
arrive at a superspace description of AdS$_4$ HS gauge theory
which is equivalent to the formulation in ordinary spacetime. This
equivalence exploits a basic property of integrable systems which
ensures that adding extra coordinates does not affect the basic
dynamics described by the system, as we shall discuss at the end
of Section 2.1.

The structure of the paper is as follows: In Section 2, we
formulate the minimal $hs(1|4)$ theory in the product of $\cN=1$,
$D=4$ superspace with the internal twistor space. Furthermore, we
eliminate the twistor variable using a curvature expansion. In
Section 3, we obtain the superspace constraints on the physical
superfields, and in particular show that they yield the correct
spectrum of the theory. In Section 4, we summarize our results,
and comment on how to extend the formalism to $\cN\geq 2$ theories
in $D=4$ and the $\cN=4$ HS gauge theory in $D=5$.


\scs{The $hs(1|4)$ Gauge Theory in Superspace}


In this section we first introduce the basic properties of the
minimal $\cN=1$, $D=4$ HS gauge theory, namely the underlying HS
symmetry algebra $hs(1|4)$ and its massless spectrum
\cite{Konshtein:1989yg,Engquist:2002vr}. Next we give the
constraints which describe the full equations of motion in
superspace, and give a formal argument for their equivalence to
the formulation in ordinary spacetime. We then generalize the
covariant curvature expansion of the constraints
\cite{Vasiliev:1990yr,Sezgin:2002ru} to superspace.

The formulation of higher spin dynamics in ordinary 4D spacetime
follows from constraints on certain one-form and zero-form master
fields which live in the product of spacetime with an internal
twistor space \cite{Vasiliev:1991vu} (our starting point is
summarized in \cite{Engquist:2002vr}). A key property of these
constraints is that all curvature components with at least one
spacetime direction are set equal to zero. Hence the form of the
constraints do not depend on the details of the spacetime
manifold. In fact, this implies that the constraints equivalently
can be formulated on extended spacetime manifolds with extra
coordinates, such as superspace.


\scss{The Closed Form of the Constraints}


The minimal $\cN=1$, $D=4$ HS theory is based on the HS algebra
$hs(1|4)$ whose maximal finite-dimensional subalgebra is
$OSp(1|4)$. The spectrum of massless physical fields of the
$hs(1|4)$ gauge theory is given by the symmetric product of two
$OSp(1|4)$ singletons, which is given in Table \ref{n1s}. The
spectrum is an UIR of $hs(1|4)$ and decomposes into a tower of
$OSp(1|4)$ multiplets labelled by a level index

\be (\ell,j)=\mx{\{}{ll}{\ell=0,1,2,\dots&\mbox{for $j=0$}\\[5pt]
\ell=-1,0,1,\dots&\mbox{for $j=1/2$}}{.} ,\la{specellj}\ee

with maximal spin

\be s_{\rm max}=2\ell+2+j\ .\ee

In particular, the spectrum contains a scalar multiplet at level
$(-1,1/2)$ and a supergravity multiplet at level $(0,0)$.

To obtain an $\cN=1$ superspace formulation of the $hs(1|4)$
theory we start from the formulation in ordinary 4D spacetime
\cite{Vasiliev:1991vu,Engquist:2002vr} and make everywhere the
replacement

\be  x^m\ra X^M=(x^m,\th^\mu,\bar \th^{\dot \mu} )\ ,\qquad
m=0,\dots,3\ ,\quad \m,\dot \m=1,2\ .\ee

The resulting theory is described in terms of a master one-form

\be \hA=dX^M\widehat A_M(X,Z;Y,\xi,\eta) +dZ^{\una} \widehat
A_{\una}(X,Z;Y,\xi,\eta)\ ,\la{hA}\ee

and master zero-form

\be\hF=\widehat \F(X,Z;Y,\xi,\eta)\ ,\ee

where $Z$ stands for $Z^{\una}=(z^\a,\zb^{\ad})$ and $Y$ for
$Y^{\una}=(y^\a,\yb^{\ad})$, which are bosonic oscillators that
generate the following associative algebra of Weyl-ordered
functions:

\be\nn \widehat f(Z;Y)\star \widehat g(Z;Y)&=& \widehat
f(Z;Y)\exp\left(i\overleftarrow{\partial}^{(+)\a}
\overrightarrow{\partial}^{(-)}_{\a}+i\overleftarrow{\partial}^{(-)\ad}
\overrightarrow{\partial}^{(+)}_{\ad}\right) \widehat g(Z;Y)\ , \\[10pt]
\partial^{(\pm)}_\a&=&\partial/\partial z^\a\pm
\partial/\partial y^\a\ .\ee

The quantities $\x$ and $\eta$ are fermionic oscillators obeying

\be \x\star\x=1\ ,\quad \x\star\eta=\x\eta=-\eta\x=\eta\star\x\
.\quad \eta\star\eta=1\ .\ee

The master fields have $(Z;Y,\x,\eta)$ expansions obeying the
following conditions

\be \tau(\widehat A)&=&-\widehat A\ ,\qquad~~ \widehat
A^{\dagger}~=~-\widehat A\ , ~\,\qquad\quad \e(\hA)~=~0 \ , \la{taua}\\[10pt]
\tau(\widehat \F)&=&\bp(\widehat\F)\ ,\quad\quad
\widehat\F^{\dagger}~=~\pi(\widehat\F)\star\C\ ,\qquad
\e(\hF)~=~0\ ,\la{tauf}\ee

where $\C=i\x\eta$, $\e$ denotes the Grassmann parity, and $\tau$,
$\pi$ and $\pb$ are maps that commute with the exterior derivative
and act on functions of $X$ and the oscillators as follows

\be \tau(\widehat f(X,z,\zb;y,\yb,\xi,\eta))&\equiv& \widehat
f(X,-iz,-i\zb;iy,i\yb,i\xi,-i\eta)\ , \\
\pi(\widehat f(X,z,\zb;y,\yb,\xi,\eta))&\equiv& \widehat
f(X,-z,\zb;-y,\yb,\xi,\eta)\ , \\
\pb(\widehat f(X,z,\zb;y,\yb,\xi,\eta))&\equiv& \widehat
f(X,z,-\zb;y,-\yb,\xi,\eta)\ . \ee

The $hs(1|4)$-valued connection and its quasi-adjoint
representation are obtained by setting $Z=0$ in \eq{taua} and
\eq{tauf}, respectively.

We propose that the full $\cN=1$ superspace formulation of the
$hs(1|4)$ theory is given by the following curvature constraints:

\be \la{eq:scon1} {\widehat D}\Fh&=&0\ ,\\[10pt]
\la{eq:scon2} {\widehat F}&=& \ft{i}4dz^{\a}\wedge dz_{\a}{\cal
V}(\Fh\star\kappa\C)+ \ft{i}4d\zb^{\ad}\wedge
d\zb_{\ad}\overline{{\cal V}}(\Fh\star\kappa^{\dagger})\ , \ee

where

\be  {\widehat D}\Fh&\equiv&d\Fh+\Fh\star\bp(\Ah)-\Ah\star\Fh\ ,\\[10pt]
{\widehat F}&\equiv&d \widehat A+ \widehat A\star  \widehat A\ ,
\la{master1}\ee

and

\be d=dX^M\partial_M + dZ^{\una}\partial_{\una}\ .\ee

We stress that the pull-back of the constraints \eq{eq:scon1} and
\eq{eq:scon2} from $(X,Z)$-space to $(x,Z)$-space yields back the
original formulation of the $hs(1|4)$ theory
\cite{Vasiliev:1991vu,Engquist:2002vr} (there is a change of sign
in $\cV(\hF\star \k)$, and hence in $\hF$, due the change between
left and right acting exterior derivatives).

The basic properties of \eq{eq:scon1} and \eq{eq:scon2} are that
they are integrable and consistent with the $\t$ and reality
conditions on the master fields given in \eq{taua} and \eq{tauf}.
The integrability of the constraints implies their invariance
under the following gauge transformations:

\be \d \Ah=d\widehat\e+\widehat\e\star\Ah-\Ah\star\widehat\e\
,\quad \d\Fh=\widehat\e\star\Fh-\Fh\star\bar\pi(\widehat\e)\
,\la{master2}\ee

where the gauge parameter $\widehat\e(X,Z;Y,\x,\y)$ obeys
\eq{taua}. A superdiffeomorphism with parameter
$\x=\x^M(X)\partial_M$ is equivalent to a field dependent gauge
transformations with parameter $\widehat\e=i_\x \widehat A$.

By fixing a gauge and making use of a subset of the constraints
one can eliminate the $Z$-dependence and thus reduce the
constraints to an equivalent but smaller set of constraints in
superspace. This yields ordinary superspace constraints, from
which one can obtain the $\theta$-expansions of the master fields,
as we shall show in Section 3. It is also possible to eliminate
the $X$ dependence, and thus obtain an equivalent formulation
entirely in $Z$-space \cite{Vasiliev:1991bu}. This yields a
non-standard form of the dynamical field equations, that could be
a convenient framework for obtaining exact solutions.

The equivalences between the formulations in superspace, ordinary
spacetime and the $Z$-space, follow from the integrability which
implies that given $\hF(0,0;Y,\x,\y)$ it is possible to determine
$\hF$ and $\hA$ for finite $X$ and $Z$ by making use of the
constraints and by fixing a gauge. Hence, the formulations in
superspace, ordinary spacetime as well as in $Z$-space are
formulations on different slices of the full $(X,Z)$-space
obtained by various (partial) gauge fixings.

From the ordinary spacetime point of view, the quantity
$\hF(0,0;Y,\x,\y)$ describes all the dynamical fields as well as
all of their spacetime derivatives that are non-vanishing on-shell
\cite{Vasiliev:1991bu}. In going to the superspace formulation,
the same quantity describes all the dynamical superfields as well
as all of their superderivatives that are non-vanishing on-shell,
implying that the two formulations are equivalent. As we shall
show, there will be one independent superfield for each multiplet
in the spectrum listed in Table \ref{n1s}.


\scss{Curvature Expansion of the Constraints}


The $Z$-dependence of the master fields can be determined in terms
of

\be \F\equiv \widehat \F{\Big |}_{Z=0}\ ,\qquad A_M\equiv \widehat
A_M{\Big |}_{Z=0}\ ,\ee

by integrating the components of \eq{eq:scon1} and \eq{eq:scon2}
that have at least one $dz^\a$ or $d\zb^{\ad}$ component. The
solution can be given as a curvature expansion in powers of $\F$ :

\be \la{eq:ce1} \widehat{\Phi}&=&\sum_{i=1}^\infty
\widehat{\Phi}^{(i)}\ ,\qquad
 \widehat{\Phi}^{(i)}{\Big |}_{Z=0}=\mx{\{}{ll}{\Phi\ ,&i=1\\0\
,&i=2,3,\ldots}{.}\ ,\\[10pt]
\la{eq:ce2} \widehat{A}_M&=&\sum_{i=0}^\infty \widehat{A}_M^{(i)}\
,\qquad \widehat{A}_M^{(i)}{\Big |}_{Z=0} =
\mx{\{}{ll}{A_M\ ,&i=0\\0\ ,&i=1,2,3,\ldots}{.}\ ,\\[10pt]
\la{eq:ce3} \widehat{A}_\a&=& \sum_{i=0}^\infty
\widehat{A}_\a^{(i)}\ , \ee

where $\widehat{\Phi}^{(i)}$, $\widehat{A}^{(i)}_\a$ and
$\widehat{A}_M^{(i)}$ are $i$th order in $\Phi$. The iterative
formulae for $\widehat{\Phi}^{(i)}$, $\widehat{A}^{(i)}_\a$ and
$\widehat{A}_M ^{(i)}$ are direct generalizations of those given
in \cite{Sezgin:2002ru} in the case of ordinary spacetime. In
particular, the superspace one-form is given by

\be \widehat{A}_M ={1\over
1+\widehat{L}^{(1)}+\widehat{L}^{(2)}+\widehat{L}^{(3)}+\cdots}
~A_M\ ,\la{hataM}\ee

where the linear operators $\widehat{L}^{(i)}$ are given in
\cite{Sezgin:2002ru}. The operator $\widehat L^{(1)}$, which will
be needed below, takes the form

\be \widehat{L}^{(1)}(\widehat f)&=&{1\over 2} \int_0^1{t'dt'
dt\over
t}\left(\left[(\Phi(-t'z,\yb,\x,\eta)\k(t'z,y)z^\a)\star\C~,~{\partial
\widehat f\over\partial
y^\a}\right]_\star\right.\nn\\&&~~~~~~~~~~~~~~~~~~~
\left.-\left[\Phi(y,t'\zb,\x,\eta)\bar\k(t'\zb,\yb)\zb^{\ad}~,~{\partial
\widehat f\over\partial \yb^{\ad}}\right]_\star\right)_{Z\ra
tZ}\hspace{-20pt}\ .\la{l1}\ee

Upon inserting the expressions for $\widehat{\F}$ and
$\widehat{A}_M$ in terms of $\F$ and $A_M$ into the remaining
components of \eq{eq:scon1} and \eq{eq:scon2}, i.e.
$\widehat{D}_M\widehat{\F}=0$ and $\widehat{F}_{MN}=0$, we obtain
curvature constraints in $\cN=1$ superspace. To analyze these
constraints we need to convert the curved index $M$ into the flat
Lorentz index $A$, which can be decomposed covariantly into
Lorentz vector and spinor indices. In order to do this, one first
identifies the exact form of the local Lorentz structure group
\cite{Vasiliev:1999ba}. Following \cite{Sezgin:2002ru}, we
decompose $A_M$ as follows:

\be A_M&=&E_M+\O_M+W_M +K_M\ ,\la{splitAM}\ee

where $W_M$ contains the higher spin gauge fields (see \eq{w}
below), and \footnote{The notation here differs from that in
\cite{Sezgin:2002ru} where $E$ denotes the $SO(3,2)$ valued gauge
field, $E=e+\o$ where $e$ is the vierbein and $\o$ the Lorentz
connection, and $\O$ denotes the rigid $SO(3,2)$ gauge field in
the AdS$_4$ background.}

\be E_M&=&  i(E_M{}^{\a}Q_\a+E_M{}^{\ad}Q_{\ad} +
E_M{}^{\a\ad}P_{\a\ad})\ ,\\[10pt]
\O_M&=& \ft{i}4(\O_M{}^{\a\b}M_{\a\b}+\O_M{}^{\ad\bd} M_{\ad\bd})\
.\ee

Here $\O_M$ is the spin connection and $E_M$ define the
supervielbein

\be E_M{}^A&=&(E_M{}^a,E_M{}^\a,E_M{}^{\ad})\ .\ee

In \eq{splitAM} we have also separated out

\be K_M=-i\,\O_M{}^{\a\b}(\Ah_\a \star \Ah_\b){\Big
|}_{Z=0}-\,{\rm h.c.}\ ,\ee

which has the effect of making the constraints manifestly Lorentz
covariant \cite{Vasiliev:1999ba} (see also \cite{Sezgin:2002ru}).
As a result the dependence of the constraints on the spin
connection is through either the covariant derivative

\be \nabla=dX^M(\partial_M+\O_M)\ , \ee

or the Riemann curvature two-form $R=d\O+\O\star\O$. We further
define the inverse supervielbein $E_A{}^M$ by $E_A{}^M
E_M{}^B=\d_A^B$ and use the notation

\be E=i dX^M E_M{}^A Q_A= i E^A Q_A\ ,\qquad
Q_A=(P_a,Q_\a,Q_{\ad})\ .\ee

The desired manifestly Lorentz invariant form of the superspace
constraints read:

\be \la{eq:hseq1} \nn \cR_{AB}+\cF_{AB}&=&-2W_{[A}\star W_{B]}
+i\sum_{\ba{c}\scriptstyle{i+j=2}\\[-5pt]\scriptstyle{i,j\geq 1}\ea}^\infty
\left(R_{AB}{}^{\a\b}(\Ah^{(i)}_\a\star\Ah^{(j)}_\b) {\Big
|}_{Z=0}+~{\rm h.c.}\right)\\ [10pt] &&-2\sum_{i+j=1}^{\infty}
\Big((i\widehat Q+\widehat W)_{[A}^{(i)}\star (i\widehat
Q+\widehat W)^{(j)}_{B]}\Big){\Big |}_{Z=0}\ , \ee \be \nn
\nabla_A\Phi+i\Phi\star\bp(Q_A)-i Q_A\star\Phi=~ -\hspace{-6pt}
\sum_{\ba{c}\scriptstyle{i+j=2}\\[-5pt] \scriptstyle{i\geq 1}\ea
}^{\infty}\hspace{-10pt}\Big(\Fh^{(i)}\star \bp((i\widehat
Q+\widehat W)_{A}^{(j)}) -(i\widehat Q+\widehat
W)_A^{(j)}\star\Fh^{(i)}\Big) {\Big |}_{Z=0}\ , \ee \vspace{-30pt}
\be \la{eq:hseq2} \ee \vspace{-25pt}

where $[AB]$ denotes graded symmetrization and we have made the
following definitions: the $OSp(1|4)$ covariant gravitational
curvature $\cR$ is defined by

\be \la{eq:adscurv} \nn\cR&\equiv& d(E+\O)+(E+\O)\star (E+\O) \\[5pt]
&=& \nabla E+d\O+\O\star\O+E\star E\nn\\[5pt]
&=& i(\cR^{\a}Q_\a +\cR^{\ad}Q_{\ad} + \cR^ {\a\ad}P_{\a\ad} +
\ft14 \cR^{\a\b} M_{\a\b} + \ft14 \cR^{\ad\bd} M_{\ad\bd})\ , \ee

with components

\be \la{cr1}\cR^\a&=&T^\a+ 2E^{\ad}\wedge E_{\ad}{}^\a\ , \\[5pt]
\la{cr2}\cR^{\a\ad}&=&T^{\a\ad}-\ft{i}2 E^\a\wedge E^{\ad}\ , \\[5pt]
\la{cr3}\cR^{\a\b}&=&R^{\a\b}-iE^\a\wedge E^\b+4 E^{\a\ad}\wedge
E_{\ad}{}^\b\  , \ee

where the superspace torsion $T^A$ and Riemann curvature
$R^{\a\b}$ are defined by

\be
T^A&=&\nabla E^A \ , \la{torsion}\\
R^{\a\b}&=&d\O^{\a\b}+ \O^{\a\c}\wedge \O_\c{}^\b\
.\la{riemann}\ee

Rigid AdS$_4$ superspace is obtained by setting $\cR=0$. The
$OSp(1|4)$ covariant higher spin curvature $\cF$ is defined by

\be \cF&=&dW+(E+\O)\star  W+W\star  (E+\O)\nn\\[10pt]
&=&\nabla W+E\star W+W\star E\ ,\ee

and has the following components

\be \cF_{AB}=2\nabla_{[A}W_{B]}+ T_{AB}{}^C W_{C} -2iQ_{[A}\star
W_{B]}-2iW_{[A}\star Q_{B]}\ ,\la{cfab}\ee

where $[AB]$ denotes graded symmetrization. Finally, the quantity
$\hA_\a$ in \eq{eq:hseq1} is defined in \eq{hA}.

The constraints \eq{eq:hseq1} and \eq{eq:hseq2} are invariant
under gauge transformations with $hs(1|4)$-valued superfield
parameters and under superspace diffeomorphisms. The analysis of
the symmetries is analogous to the case of ordinary spacetime;
see, for example, \cite{Sezgin:2002ru} for a discussion and the
explicit form of the gauge transformations.

The constraints \eq{eq:hseq1} and \eq{eq:hseq2} define the full
$hs(1|4)$ gauge theory in superspace. As explained in Section 2.1,
this superspace formulation is equivalent to the formulation in
ordinary spacetime. To verify this directly, one begins with the
restriction of \eq{eq:hseq1} and \eq{eq:hseq2} to the bosonic
submanifold, which, by construction, yields the curvature
constraints in ordinary spacetime described in
\cite{Engquist:2002vr}. Thus, in order to verify the equivalence
between the two formulations it suffices to show that the
superspace formulation does not contain any additional propagating
degrees of freedom nor any new constraints. This can be analyzed
in a manifestly superspace covariant expansion in which both $\F$
and the higher spin gauge fields $W_M$ are weak fields, while the
gauge fields of the supergravity multiplet are treated exactly
\cite{Sezgin:2002ru}. If the leading order does not yield any new
degrees of freedom or constraints, then this must hold to all
orders in the weak field expansion. Indeed this will be shown to
be the case in the next section.

{\footnotesize \tabcolsep=1mm
\begin{table}[t]
\bec
\begin{tabular}{|c|cccccccccccl|}\hline
& & & & & & & & & & & & \\
{\large${}_{{(\ell,j)}}\backslash s$} & $0$ & \ns{$\ft12$} & $1$ &
\ns{$\ft32$} & $2$ & \ns{$\ft52$} & $3$ & \ns{$\ft72$} & $4$ &
\ns{$\ft92$} & $5$ & $\cdots$
\\
& & & & & & & & & & & & \\
\hline
& & & & & & & & & & & & \\
$(-1,1/2)$ & $1\!+\!{\bar 1}$ & $1$ & & & & & &
& & & & \\
$(0,0)$ & & & & $1$ & $1$ & & & & &
& & \\
$(0,1/2)$ & & & & & $1$ & $1$ & & & &
& & \\
$(1,0)$ & & & & & & & & $1$ & $1$ &
& & \\
$(1,1/2)$ & & & & & & & & & $1$ & $1$
& & \\
$\vdots$ & & & & & & & & & & & &
 \\ \hline
\end{tabular}
\ec \caption{{\small  The spectrum of massless physical fields of
the minimal $\cN=1$, $D=4$ higher spin gauge theory arranged into
levels of $\cN=1$ supermultiplets labelled by $(\ell,j)$ and with
$s_{\rm max}=2\ell+2+j$.}} \label{n1s}
\end{table}}


\scs{Analysis of Superspace Constraints}


In this section we analyze the superspace constraints
\eq{eq:hseq1} and \eq{eq:hseq2} in the first order in the weak
field expansion, in which we treat $\F$ and $W_M$ as weak fields
while the supervielbein $E_M$ contains information of the full
ordinary $\cN=1$ supergravity theory (without higher derivative
corrections). Indeed, we will find the superspace constraints that
describe the ordinary on-shell $\cN=1$, $D=4$ supergravity with
cosmological constant. Moreover, we will find the constraints on
the superfields describing the scalar and the higher spin
multiplets in the first order in the weak fields.


\scss{Weak Field Expansion}


Assuming that both $\F$ and the higher spin gauge fields $W_M$ are
weak fields, the leading contributions to the superspace
constraints \eq{eq:hseq1} and \eq{eq:hseq2} are

\be \cR+\cF^{(1)}&=&-\{E,\widehat E^{(1)}\}_{\star} {\Big
|}_{Z=0}=\{E,\widehat L^{(1)}(E)\}_{\star} {\Big |}_{Z=0}\
,\la{lc1}\\[8pt]&=&-\left( E^{\ad}\wedge
E_{\ad}{}^{\a}\fr{\partial}{\partial y^{\a}}
\fr{\partial}{\partial \xi}+iE^{\a\ad}\wedge
E_{\ad}{}^{\b}\fr{\partial}{\partial y^{\a}}
\fr{\partial}{\partial
y^{\b}}\right)\F(y,0,\xi,\eta)\nn\\[5pt]&&-
\left(E^{\a}\wedge E_{\a}{}^{\ad}\fr{\partial}{\partial \yb^{\ad}}
\fr{\partial}{\partial \xi}+iE^{\ad\a}\wedge
E_{\a}{}^{\bd}\fr{\partial}{\partial \yb^{\ad}}
\fr{\partial}{\partial
\yb^{\bd}}\right)(\F(0,\yb,\xi,\eta)\star\C)\ ,\nn\ee

and

\be \la{lc2} \nabla_A\Phi+i\Phi\star\bp(Q_A)-iQ_A\star\Phi&=&0\
,\ee

where the higher spin field strength $\cF^{(1)}$ is given by
\eq{cfab} with $T_{AB}{}^C$ set equal to the rigid AdS$_4$
superspace torsion and $\widehat L^{(1)}$ is defined in \eq{l1}.
The supervielbein and the spin connection describe a curved
superspace, which as we shall see is pure supergravity in the
leading order. The integrability of \eq{lc1} and \eq{lc2} holds
modulo terms which involve the supergravity field strengths, which
are given in \eq{s32} and \eq{s2} below, times other weak fields.

We proceed by decomposing $W_M$ and $\F$ into levels as follows
\cite{Engquist:2002vr}

\be\la{w}W_M&=&W^{(0,1/2)}_M+\sum_{\ell\geq 1}(W^{(\ell,0)}_M
+W^{(\ell,1/2)}_M)\ ,\\[10pt]
\F&=&\F^{(-1,1/2)}+\sum_{\ell\geq
0}(\F^{(\ell,0)}+\F^{(\ell,1/2)})\ ,\la{philevel1}\ee

where

\be \la{wellj} W_M^{(\ell,j)}&=&\sum_{p+q+r=4\ell+2+2j}
W^{(\ell,j)}_{M,r}(p,q)\x^r\eta^{2j}\ , \\[10pt]
\F^{(\ell,j)}&=&
C^{(\ell,j)}+\pi\Big((C^{(\ell,j)})^\dagger\Big)\star\C\
,\la{philevel2}\\[10pt]
C^{(\ell,j)}&=&\sum_{\ba{c}\scriptstyle{q-p-r}\\[-5pt]\scriptstyle{=4\ell+3+2j}\ea}
\F^{(\ell,j)}_{r}(p,q)\x^r\eta^{1-2j}\ .\la{philevel3}\ee

and $(p,q)$ refers to the $y$ and $\yb$ expansion as defined in
\eq{app1}. Since the adjoint and quasi-adjoint actions of $Q_A$ on
$W^{(\ell,j)}$ and $\F^{(\ell,j)}$ preserve the level index, it
follows that the constraints \eq{lc1} and \eq{lc2} split into a
separate set of constraints for each level:

\be \nn\cR &=&   -\left(E^{\ad}\wedge
E_{\ad}{}^{\a}\fr{\partial}{\partial y^{\a}}
\fr{\partial}{\partial \xi}+iE^{\a\ad}\wedge
E_{\ad}{}^{\b}\fr{\partial}{\partial y^{\a}}
\fr{\partial}{\partial y^{\b}}\right)\F^{(0,0)}(y,0,\xi,\eta)-{\rm
h.c.}\ , \\ \la{graveq} \\[10pt]
\nn \cF^{(1)(\ell,j)}&=&  -\left(E^{\ad}\wedge
E_{\ad}{}^{\a}\fr{\partial}{\partial y^{\a}}
\fr{\partial}{\partial \xi}+i E^{\a\ad}\wedge
E_{\ad}{}^{\b}\fr{\partial}{\partial y^{\a}}
\fr{\partial}{\partial
y^{\b}}\right)\F^{(\ell,j)}(y,0,\xi,\eta)-{\rm h.c.}\ , \\
\la{lfeq}\ee
\be \nabla_A\Phi^{(\ell,j)}+i
\Phi^{(\ell,j)}\star\pb(Q_A)-iQ_A\star\Phi^{(\ell,j)}&=&0\
,\la{lphieq} \ee

where the level decomposition of the higher spin field strength is
given by

\be \cF&=&\cF^{(0,1/2)}+
\sum_{\ell=1}^\infty(\cF^{(\ell,0)}+\cF^{(\ell,1/2)})\ ,\\
\cF^{(\ell,j)}&=&\nabla W^{(\ell,j)}+E\star
W^{(\ell,j)}+W^{(\ell,j)}\star E\la{fellj}\ee

and $\cF^{(\ell,j)}$ has the same expansion in $(y,\yb,\x,\eta)$
as in \eq{wellj}.


\scss{The $\cR$-constraint}


In this section we show that the $OSp(1|4)$ valued curvature $\cR$
subject to the constraint \eq{graveq} describes the supergravity
multiplet residing at level $(0,0)$.

From \eq{graveq} and eqs.~\eq{cr1}--\eq{cr3} it follows that up to
first order in the weak field expansion the superspace torsion and
Riemann tensor are given by

\be T^{\a\ad}&=&\ft{i}2 E^{\a}\wedge E^{\ad}\ ,\la{a2} \\[10pt]
T^{\a}&=& - 2E^{\ad}\wedge E_{\ad}{}^{\a} -2 E^{\b\ad}\wedge
E_{\ad}{}^\c\Psi^\a{}_{\b\c}\ , \la{a1}\\[10pt]
R^{\a\b}&=&i E^{\a}\wedge E^{\b}+2i E^{\ad}\wedge
E_{\ad}{}^\c\Psi^{\a\b}{}_\c-4 E^{\a\ad}\wedge
E_{\ad}{}^{\b}-2E^{\c\ad}\wedge E_{\ad}{}^\d C^{\a\b}{}_{\c\d}
,\qquad\la{a3}\ee

where the gravitino curvature and Weyl tensor are defined by

\be \Psi_{\a\b\c}&\equiv& {\partial^3\over\partial y^\a\partial
y^\b\partial y^\c}{\partial\over\partial
\x}\F{\Big |}_{Y=\x=\eta=0}\ ,\la{s32}\\[10pt]
C_{\a\b\c\d}&\equiv&{\partial^4\over\partial y^\a\partial
y^\b\partial y^\c\partial y^\d}\F{\Big |}_{Y=\x=\eta=0}\
.\la{s2}\ee

The Weyl tensor is the spinor derivative of the gravitino
curvature, as given in \eq{fsc2}. The constraints \eq{a2}--\eq{a3}
describe the on-shell $\cN=1$ pure supergravity multiplet with
cosmological constant. To see this, we note that \eq{a2} and
\eq{a1} contain the constraints:

\be T_{\underline{\a\b}}{}^c=-i(\C^c)_{\underline{\a\b}}\ ,\qquad
T_{Ab}{}^c=0\ ,\ee

where $\underline{\a}=(\a,\ad)$, which describe the $\cN=1$
off-shell supergravity multiplet in superspace with auxiliary
fields given by a complex scalar field $S+iP$ and a real vector
\cite{Grimm:1979ch} (see also \cite{Bagger:1990qh}). The remaining
constraints in \eq{a2}--\eq{a3} amount to setting the auxiliary
vector equal to zero and the pseudo-scalar $P$ equal to a constant
 (in units where the AdS$_4$ radius is set equal
to $1$).


\scss{The $\cF^{(1)}$-constraint}


In this section we analyze the constraint \eq{lfeq} and show that
their $\th$-expansion contains the higher spin multiplets.

The $AB=\a\b$ and $AB=a\b$ components of \eq{lfeq} read

\be \cF^{(1)}_{\underline{\a\b}}\equiv
2\nabla_{(\underline{\a\phantom{\b}\!\!\!}}W_{\underline{\b})}-i
(\C^c)_{\underline{\a\phantom{\b}\!\!\!}\underline{\b}}W_c
-2i\{Q_{(\underline{\a\phantom{\b}\!\!\!
}},W_{\underline{\b})}\}_{\star}=0\ ,\la{alfabeta}\ee

and

\be \nn \cF^{(1)}_{a\b}&\equiv& \nabla_a W_\b-\nabla_\b W_a-\ft12
(\s_a)_\b{}^{\cd} W_{\cd} - i[P_a,W_\b]_{\star}
+ i[Q_\b,W_a]_{\star} \\[10pt]&=& -\fr{1}{4}(\s_a)_\b{}^{\cd}
{\partial\over \partial \yb^{\cd}} {\partial\over\partial\x}
\F(0,\yb,\x,\eta) \star \C\ ,\la{aalfa}\ee

where the level index $(\ell,j)$ has been suppressed. The
structure of these constraints is similar to that of ordinary
Yang-Mills theory in $\cN=1$ superspace, and we may proceed in an
analogous fashion in order to show that no new component fields
arise upon $\theta$-expanding $W_\a$ and $W_a$.

First, from $\delta W_\a=\nabla_\a \e$, where $\e$ is the
$hs(1|4)$ valued gauge parameter, it follows that we can fix a
gauge in which

\be W_\a|=0\ .\la{walfa}\ee

From \eq{alfabeta} and \eq{walfa} it follows that

\be \nabla_{(\underline{\a}} W_{\underline{\b})}|=
\ft{i}2(\C^c)_{\underline{\a\phantom{\b}\!\!\!}\underline{\b}}W_c|\
. \la{nablaalfawbetasymm}\ee

By using the $\th^2$-component of $\e$, i.e.
$(\nabla_{\underline{\a}}\nabla_{\underline{\b}}-
\nabla_{\underline{\b}}\nabla_{\underline{\a}}) \e|$, we can
impose the further gauge condition:

\be (\nabla_{\underline{\a}}
W_{\underline{\b}}-\nabla_{\underline{\b}} W_{\underline{\a}})|=0\
.\la{nablaalfawbetaas}\ee

Hence the first non-trivial components in the $\theta$-expansion
of $W_\a$ are the leading components of $W_a$, i.e. $W_a|$. We
proceed by examining $\nabla_\a W_b|$. From \eq{aalfa} and
\eq{walfa}, \eq{nablaalfawbetasymm} and \eq{nablaalfawbetaas} it
follows that

\be \nabla_\a
W_b|=i[Q_\a,W_b|~]_\star-i\psi_b{}^{\bd}(\s^a)_{\bd\a}W_a|+\fr{1}{4}(\s_b)_\a{}^{\bd}
{\partial\over \partial \yb^{\bd}} {\partial\over\partial\x}
\F^{(\ell,j)}(0,\yb,\x,\eta){\Big |} \star \C\
,\qquad\la{nablaalfawb}\ee

where $\psi_a{}^\a=E_a{}^\a|$. In order to evaluate the last term
in \eq{nablaalfawb}, we examine the $AB=ab$ component of
\eq{lfeq}. This constraint has the same form as in the formulation
in ordinary spacetime. The latter have been analyzed in detail in
\cite{Vasiliev:1987td}. Using these results we conclude that the
last term in \eq{nablaalfawb} can be written in terms of spacetime
derivatives of $W_a|$. More specifically, the gauge fields
$W_{a,r}(p,q)$ with $|p-q|\leq 1$ are independent, while those
with $|p-q|\geq 2$ are auxiliary fields that are $[\ft12|p-q|]$
spacetime derivatives of the independent gauge fields. Moreover,
it follows that the spin $s$ field strengths contained in
$\F^{(\ell,j)}(0,\yb,\x,\eta)$ can be written as $[s]$ spacetime
derivatives of the independent gauge fields. Combining the above
results we deduce that $W_A$ can be $\theta$-expanded in terms of
$W_a|$ and its spacetime derivatives.

It is also possible to eliminate the auxiliary fields in a
manifestly superspace covariant manner. To this end, we first use
\eq{alfabeta} to solve for $W_a$ in terms of $W_\a$:

\be W_a={i\over 2} (\C_a)^{\underline{\a\b}}\left(
\nabla_{\underline{\a\phantom{\b}\!\!\!}}
W_{\underline{\b}}-i\{Q_{\underline{\a\phantom{\b}\!\!\!}},
W_{\underline{\b}}\}_\star\right)\ .\ee

Inserting this into \eq{aalfa} we then obtain

\be \nn&&i[P_a,W_\b]_\star-{i\over 2} (\C_a)^{\underline{\c\d}}
\{Q_\b,\{Q_{\underline{\c}},W_{\underline{\d\phantom{\c}\!\!\!}}\}_\star\}_\star\\[5pt]
\nn &=&\nabla_aW_\b-{i\over 2}
(\C_a)^{\underline{\c\d}}\nabla_\b\left(\nabla_{\underline{\c}}
W_{\underline{\d\phantom{\c}\!\!\!}}-i\{Q_{\underline{\c}},
W_{\underline{\d\phantom{\c}\!\!\!}}\}_\star\right) -{1\over 2}
(\C_a)^{\underline{\c\d}}\{Q_\b,\nabla_{\underline{\c}}
W_{\underline{\d\phantom{\c}\!\!\!}}\}_\star\\[5pt]&&-{1\over
2}(\s_a)_\b{}^{\cd}W_{\cd}+ \fr1{4}(\s_a)_\b{}^{\cd}
{\partial\over \partial \yb^{\cd}} {\partial\over\partial\x}
\F^{(\ell,j)}(0,\yb,\x,\eta) \star \C\ .\ee

Expanding in $y$ and $\yb$, the $(m,n)$ component of the above
equations are supposed to be solved for $W_\b(m+1,n-1)$ in terms
of $W_\b(m',n')$ with $|m'-n'|< |m-n|$ covered by various
combinations of spinor and vector derivatives (the structure is
similar to that of the $AB=ab$ components of \eq{lfeq}). Hence a
subset of the $W_\a(m,n)$ are independent gauge superfields, and
the remaining $W_\a(m,n)$ and all of $W_a$ are auxiliary
superfields. The details of this, and in particular, the
constraints obeyed by the independent gauge superfields, remain to
be worked out.


\scss{The $\F$-constraint}


In this section we show that the master field $\F$ subject to the
constraint \eq{lphieq} contains the physical scalar multiplet at
level $(-1,1/2)$ and the field strengths (i.e. the Weyl tensors)
of the higher level physical gauge fields. The $A=a$ component of
\eq{lphieq} has the same structure as in the formulation in
ordinary spacetime. Hence the components $\F^{(\ell,j)}_r(k,k+q)$
with $k>0$ are auxiliary fields which can be expressed as $k$
bosonic derivatives of the components $\F^{(\ell,j)}_r(0,q)$.

\begin{center} {\it The field strength multiplets}\end{center}

Eq.~\eq{lfeq} identifies $\F^{(\ell,j)}_r(0,q)$ for $\ell\geq0$,
$j=0,1/2$, $q=4\ell+3+2j+r$ with the non-vanishing spin $s=q/2$
field strengths discussed in Section 3.2 and 3.3.

It is remarkable that \eq{lphieq} also yields the on-shell
constraints on the superfield $\F^{(\ell,j)}_{0\,\ad_1\dots
\ad_{2s}}$ ($s=2\ell+3/2+j$) whose $\th$-expansion yields the
level $(\ell,j)$ field strength supermultiplet. We find these
constraints to be:

\be \nabla_\b \F^{(\ell,j)}_{0\,\ad_1\dots \ad_{2s}}&=&0\
,\la{fsc1}\\[10pt]
\nabla_{\bd}\F^{(\ell,j)}_{0\,\ad_1\dots
\ad_{2s}}&=& (-1)^{2s}\F^{(\ell,j)}_{1\,\bd\ad_1\dots\ad_{2s}}\ ,\la{fsc2}\\[10pt]
\nabla_{\bd}\F^{(\ell,j)}_{1\,\ad_1\dots
\ad_{2s+1}}&=&-i(-1)^{2s}(2s+1)\e_{\bd(\ad_1}\F^{(\ell,j)}_{0\,\ad_2\dots\ad_{2s+1})}\
,\la{fsc3}\ee

where\\[-40pt]

\be \F^{(\ell,j)}_{r\,\ad_1\dots \ad_{q}}={\partial^q\over\partial
\yb^{\ad_1}\cdots\partial \yb^{\ad_q}}\F^{(\ell,j)}_r(0,q)|_{Y=0}\
.\ee

The independent superfield is evidently
$\F^{(\ell,j)}_{0\,\ad_1\dots \ad_{2s}}$. Indeed we can eliminate
$\F^{(\ell,j)}_{1\,\ad_1\dots \ad_{2s+1}}$ from the symmetric part
of \eq{fsc2}. From \eq{fsc2} and \eq{fsc3} we then find

\be \nabla^{\bd} \F^{(\ell,j)}_{0\,\bd\ad_2\dots \ad_{2s}}&=&0\
,\la{fsc4}\\[10pt]\nabla_{\bd}\nabla_{\phantom{\bd}\!\!\!\!\!(\ad_1}
\F^{(\ell,j)}_{0\,\ad_2\dots
\ad_{2s+1})}&=&-i(2s+1)\e_{\bd(\ad_1}\F^{(\ell,j)}_{0\,\ad_2\dots
\ad_{2s+1})}\ .\la{fsc5}\ee

Hence the complete set of constraints on
$\F^{(\ell,j)}_{0\,\ad_1\dots \ad_{2s}}$ in the first order in the
weak field expansion are \eq{fsc1}, \eq{fsc4} and \eq{fsc5}.


\begin{center} {\it The level $(-1,1/2)$ scalar multiplet}\end{center}


It remains to analyze $\F^{(-1,1/2)}_0(0,0)$ and
$\F^{(-1,1/2)}_1(0,1)$ which contain the superfields

\be \superphi\equiv \F{\Big |}_{Y=\x=\eta=0}\ ,\qquad
\Psi_{\ad}\equiv
-{\partial\over\partial\x}{\partial\over\partial\yb^{\ad}} \F{\Big
|}_{Y=\x=\eta=0}\ .\la{equivf}\ee

From the $A=a$ components of \eq{lphieq} it follows that the
leading components $\phi=\superphi|$ and $\psi_{\ad}=\Psi_{\ad}|$
obey physical field equations with masses $m^2_\phi=-2$ and
$m_\psi=0$. The $A=\a$ and $A=\ad$ components of \eq{lphieq}
yields

\be \nabla_\a\superphi=0\ ,\qquad
\nabla_{\ad}\superphi=\Psi_{\ad}\ ,\quad
\nabla_{\bd}\nabla_{\ad}\superphi= -i\e_{\bd\ad} \superphi\
,\la{chiral}\ee

which are the appropriate constraints for the on-shell scalar
multiplet. Note that, in the presence of cosmological constant,
the quantity $\nabla_{\ad}\nabla_{\bd}\superphi$, where
$\superphi$ is chiral, can contain both $\superphi$ and
$\superphi^\dagger$ terms.

In summary, the only physical superfield that arises in $\F$ is
$\superphi$, which describes the scalar multiplet residing at
level $(-1,1/2)$ in Table 1.


\scs{Comments}


In this paper we have given the $\cN=1$ superspace formulation of
the massless 4D higher spin gauge theory based on the minimal
$\cN=1$ higher spin algebra $hs(1|4)$. The general arguments for
the equivalence between the superspace formulation and the
formulation in ordinary spacetime are given at the end of Section
2.1, and explicitly verified  in Section 3.

In the first order in the weak field expansion (see Section 3.1),
the resulting on-shell constraints on the supergravity multiplet
are given in \eq{a2}, \eq{a1} and \eq{a3}, those on the higher
spin field strengths in \eq{fsc1}, \eq{fsc4} and \eq{fsc5}, and
those on the scalar multiplet in \eq{chiral}. The supergravity
constraints describe the pure supergravity field equations without
higher derivative corrections and with cosmological constant. The
corrections to these equations, including the stress-energy from
the scalar and higher spin multiplets as well as higher order
curvature corrections, can be obtained order by order in the
covariant weak field expansion scheme, as discussed for example in
\cite{Sezgin:2002ru}.

The master constraints \eq{eq:scon1} and \eq{eq:scon2} generalize
straightforwardly to the supersymmetric higher spin gauge theories
based on the $\cN\geq 2$ extended higher spin algebras discussed
in \cite{Engquist:2002vr}. In all these algebras the scalar master
field $\hF$ obeys a reality condition of the form
$\hF^\dagger=\pi(\hF)\star\C$, where $\C$ is a suitable operator
formed out of the fermionic oscillators used in the construction
of the extended higher spin algebras. The equivalence between the
resulting superspace formulations and the previous formulations in
bosonic spacetime follows from the general argument given at the
end of Section 2.1.

It is well-known that the $\cN=8$ supergravities in $D=4$ exhibit
hidden symmetries. In the superspace formulation, these symmetries
arise naturally as a result of embedding the $28$ vector fields
into the composite $SU(8)$ connection in superspace
\cite{Brink:1979nt,Howe:1982tp}. The required torsion constraints
also arise in the leading order in the weak field expansion of the
$\cN=8$ higher spin gauge theory, as a subset of the constraints
on the $OSp(8|4)$ valued curvature $\cR$ (see Section 3.2 for the
analogous discussion in the case of $\cN=1$). There are, of
course, corrections to the torsion constraints, and it would be
interesting to examine to what extent they affect the potential
for the $70$ supergravity scalars, $\f_{ijkl}$. In principle, this
potential can be computed directly by evaluating the scalar field
equation order by order by in the curvature expansion scheme
\cite{Sezgin:2002ru}. However, this procedure does not rely on the
superspace formulation. To make use of the superspace formulation,
one should first find the manifestly $SL(2,{\mathbb C})\times
SU(8)$ covariant formulation, i.e. generalize the identification
of the exact spin connection $\O_M$ in \eq{splitAM} to include
also the $SU(8)$-connection, and then examine the resulting
corrected supergravity torsion constraints, in which terms
depending on the scalars $\f_{ijkl}$  (but not their derivatives)
can give rise to corrections to the potential. Modulo such
corrections, the potential for $\f_{ijkl}$ in the higher spin
gauge theory is given by the scalar potential of the supergravity
theory.

It is straightforward to generalize the formalism described in
this paper to obtain an $\cN=4$, $D=5$ superspace formulation of
the 5D HS gauge theory based on the $hs(2,2|4)$ algebra
\cite{Sezgin:2001yf}. The massless spectrum consists of the
supergravity multiplet and a tower of $s_{\rm max}=4,6,8,\ldots$
multiplets with spin range $4$. The superspace master fields are a
zero-form $\F$ and a one-form $A_M$ which are expansions in terms
of bosonic oscillators $y_\a$ and $\yb_\a$ ($\a=1,\dots,4$) and
fermionic $SU(4)$ oscillators $\x^i$ and $\bar \x_i$
($i=1,\dots,4$) governed by the same $\t$ and hermicity conditions
as in \cite{Sezgin:2001yf}. The supertranslations,
$Q_A=(P_a,Q_\a^i,Q_{\a i})$ are given by
$P_a=(\C_a)^{\a\b}y_\a\yb_\b$, $Q^i_\a=\x^i\yb_\a$ and $Q_{\a
i}=\bar \x_i y_\a$. Applying the superspace formalism described in
this paper yields the following linearized $\F$-constraint:

\be \nabla_A\F+Q_A\star \F-\F\star \pi(Q_A)=0\ ,\la{5d}\ee

where $\nabla_A$ is the covariant derivative in the rigid AdS$_5$
superspace obeying $d\O+\O\star \O=0$ and $\pi$ is defined in
\cite{Sezgin:2001yf}. The above constraint is integrable and
consistent with the $(y,\yb,\x,\bar \x)$-expansion of $\F$. As
shown in \cite{Sezgin:2001yf}, the $A=a$ component of \eq{5d}
yields the correct linearized field equations for the $hs(2,2|4)$
theory. We expect that the $A=\a i$ components of \eq{5d} contain
the corresponding superspace constraints.

Assuming that the $hs(2,2|4)$ theory is related to the $1/N^2$
expansion of the free $SU(N)$ SYM theory in $\cN=4$, $d=4$
superspace, the supergravity multiplet couples to the
superconformal current and the $s_{\rm max}=4,6,\ldots$ multiplets
couple to the higher spin supercurrents of the free SYM theory.
Hence, the reduction of the $\F$-constraint to $d=4$, $\cN=4$
superspace must reproduce the constrained superfield strengths of
the prepotentials coupling to the currents of the free SYM theory.

The superspace formulation of ordinary supergravity is
indispensable in their coupling to extended objects, which are
described by $\k$-symmetric brane actions. It would be interesting
to see if the superspace formulation of the HS gauge theories
presented in this paper could be utilized for their coupling to
extended objects, which in turn might provide a dual description
of the spacetime physics at large energies.

Finally, the general argument given at the end of Section 2.1
suggests further extensions of the 4D (super)spacetime by extra
coordinates, possibly infinitely many, corresponding to a suitable
coset of the HS algebra and with the property that a finite number
of independent dynamical fields contain all component fields in
the master fields in their expansion in the extra coordinates.
Free field constructions of this type have been considered in
\cite{Vasiliev:2001zy,Devchand:1998ff}.

\bigskip

\section*{\sc Acknowledgements}

E.S. would like to thank ITP in Uppsala University, and P.S. would
like to thank the George P. and Cynthia W. Mitchell Institute for
Fundamental Physics, for hospitality. The work of E.S. is
supported in part by NSF Grant PHY-0070964.

\newpage

\begin{appendix}

\scs{Notations and Conventions}

We use the following notation for expanding functions of the
oscillators $y$ and $\yb$:

\be f(y,\yb)&=&\sum_{m,n=0}^\infty f(m,n)\ ,\nn\\
f(m,n)&=&{1\over m!n!}f_{\a_1\dots \a_m,\ad_1\dots
\ad_n}y^{\a_1}\cdots y^{\a_m} \yb^{\ad_1}\cdots \yb^{\ad_n}\
.\la{app1}\ee

Differentiation with respect to the fermionic oscillators is
defined by

\be {\partial\widehat f\over\partial \x} ={1\over 2}[\x,\widehat
f]_\star\ .\ee

The van der Waerden symbols are defined by

\be (\s^a \bar \s^b)_{\a\b}&=&\eta^{ab}\e_{\a\b}+(\s^{ab})_{\a\b}\
,\\ \frac{1}2  \e^{abcd}\s_{cd}&=&i\s^{ab}\ ,\\(\bar
\s^{ab})_{\ad\bd}&=&((\s^{ab})_{\a\b})^\dagger\ ,\ee

where $(\bar \s^a)_{\ad\b}=
((\s^a)_{\a\bd})^\dagger=(\s^a)_{\b\ad}$,
$\e_{\a\b}\e^{\c\d}=2\d_{[\a}^\c\d_{\b]}^\d$,
$\e_{\ad\bd}=(\e_{\a\b})^\dagger$, and spinor contractions are
according to north-west-south-east rule. The completeness
relations are given by

\be\ba{lcllcl}
(\s^a)_{\a\ad}(\s_a)_{\b\bd}&=&-2\e_{\a\b}\e_{\ad\bd}\
,&(\s^{ab})_{\a\b}(\s_{ab})_{\c\d}&=&-8\e_{\a(\c}\e_{\d)\b}\ ,
\\[7pt](\s^a)^{\a\ad}(\s^b)_{\a\ad}&=&-2\eta^{ab}\ ,&
(\s^{ab})^{\a\b}(\s_{cd})_{\a\b}&=&4\d^{ab}_{cd}-2i\e^{ab}{}_{cd}\
.\ea\ee

The $OSp(1|4)$ generators in spinor basis are given by

\be Q_\a=\ft12 y_\a\x\ ,\quad M_{\a\b}=y_\a y_\b\ ,\quad
P_{\a\ad}= y_\a\yb_{\ad}\ ,\ee

and they obey

\be \ba{lcllcl}\{Q_\a,Q_\b\}&=&\ft12 M_{\a\b}\ ,\quad& \{Q_\a,
Q_{\ad}\}&=&\ft12 P_{\a\ad}\ ,\\[5pt] [Q_\a,M_{\b\c}]&=&4i\e_{\a(\b}Q_{\c)}\ ,&
[Q_\a,P_{\b\bd}]&=&2i \e_{\a\b} Q_{\bd}\ ,\\[5pt]
[M_{\a\b},M_{\c\d}]&=&4i(\e_{\a(\c}M_{\d)\b}+\e_{\b(\c}M_{\d)\a})\
,&
[P_{\a\ad},M_{\b\c}]&=&4i\e_{\a(\b}P_{\c)\ad}\ ,\\[5pt] [P_{\a\ad},P_{\b\bd}]&=&
2i(\e_{\a\b}M_{\ad\bd}+\e_{\ad\bd}M_{\a\b})\ .\ea\ee

The $SO(3,2)$ commutation relations are
$[M_{\hat{a}\hat{b}},M_{\hat{c}\hat{d}}]=-i(\eta_{\hat{b}\hat{c}}M_{\hat{a}\hat{d}}+3
\,\mbox{more})$, where $\eta_{\hat{a}\hat{b}}={\rm diag}(-+++-)$
and

\be \ba{lcllcl} M_{ab}&=&\ft18
(\s_{ab})^{\a\b}M_{\a\b}+\mbox{h.c.}\ ,\quad&
P_a&=&M_{a5}=\ft14(\s_a)^{\a\ad}P_{\a\ad}\ ,\\[5pt] M_{\a\b}&=&(\s^{ab})_{\a\b}M_{ab}\ ,&
P_{\a\ad}&=&-2(\s^a)_{\a\ad}P_a\ .\ea\ee

An $SO(3,2)$ valued element $\L$ is expanded as \vspace{-10pt}

\be \L &=&\ft12 \L^{ab}M_{ab}+\L^a P_a=\ft14
\L^{\a\b}M_{\a\b}+\mbox{h.c.}+\L^{\a\ad}P_{\a\ad}\ ,\ee

where

\be\ba{lcllcl}
\L_{ab}&=&\ft12(\s_{ab})^{\a\b}\L_{\a\b}+\mbox{h.c.}\
,\quad&\L_a&=&-2(\s_a)^{\a\ad}\L_{\a\ad}
\ ,\\[5pt]\L_{\a\b}&=&\ft14(\s^{ab})_{\a\b}\L_{ab}
\ ,&\L_{\a\ad}&=&\ft14(\s^a)_{\a\ad}\L_a\ .\ea\la{app1}\ee

We use the following superspace conventions:

\be G&=&{1\over p!}E^{A_1}\cdots E^{A_p}G_{A_p\dots A_1}=
{1\over p!}dX^{M_1}\cdots dX^{M_p} G_{M_p\dots M_1}\ ,\\
dG&=& {1\over p!}dX^{M_1}\cdots dX^{M_p}
dX^N{\partial\over\partial X^N} G_{M_p\dots M_1}\ .\ee

The right-action of the exterior derivative implies that
$d(GH)=GdH+(-1)^q (dG)H$ for a $p$-form $G$ and $q$-form $H$. The
hermitian conjugation of $GH$ is

\be (GH)^\dagger=(-1)^{pq}H^\dagger
G^\dagger=(-1)^{\e(G)\e(H)}G^\dagger H^\dagger\ .\ee

The conversion between Lorentz and spinor indices for the vector
components of a one-form $G=E^A G_A$ is given by

\be E^a&=&-2(\s^a)_{\a\ad}E^{\a\ad}\ ,\quad
G_a=\ft14(\s_a)^{\a\ad}G_{\a\ad}\ ,\\
E^{\a\ad}&=&\ft14(\s^a)_{\a\ad}E_a\ ,\quad
G_{\a\ad}=-2(\s^a)_{\a\ad}G_a\ .\ee

The Lorentz covariant derivatives are defined by:

\be \nabla V^\a&=& dV^\a+ V^\b\O_\b{}^\a\ ,\\
\nabla V^a&=&dV^a+V^b \O_b{}^a\ ,\ee

where $\O_{ab}$ and $\O_{\a\b}$ are related to each other as in
\eq{app1}. The graded commutator of two covariant derivatives is
given by

\be 2\nabla_{[A}\nabla_{B]}V^C&=&(-1)^{D(A+B)}V^D
R_{AB,D}{}^C-T_{AB}{}^D \nabla_D V^C\ ,\ee

and the torsion identity reads

\be \nabla_{[A}
T_{BC]}{}^D+T_{[AB|}{}^ET_{E|C]}{}^D=R_{[AB,C]}{}^D\ .\ee

The non-vanishing rigid AdS$_4$ superspace torsions and curvatures
are

\be T_{\a\bd}{}^c&=&-i(\s^c)_{\a\bd}\ ,\quad
T_{a\bd}{}^\c=-\ft12(\s_a)_{\bd}{}^\c\ ,\\
R_{\a\b}{}^{\c\d}&=&2i\d^\c_{(\a}\d^\d_{\b)}\ ,\quad
R_{ab}{}^{\c\d}=\ft12 (\s_{ab})^{\c\d}\ .\ee

Our conventions are such that a scalar field $\phi$ with mass $m$
and AdS energy $E$ obeys

\be (\nabla^2-m^2)\phi=0\ ,\quad E(E-3)=m^2R^2\ ,\ee

where $R$ is the AdS radius, which we have set equal to $1$.

\end{appendix}

\pagebreak





\end{document}